\documentclass[10pt,twocolumn,twoside]{IEEEtran}

\usepackage{pgf,tikz}
\usetikzlibrary{arrows}

\usepackage{amsfonts}
\usepackage{amssymb}
\usepackage{amsmath}
\usepackage{mathptmx}
\usepackage{color}
\usepackage{cancel}
\usepackage{bbm}
\usepackage{psfrag}
\usepackage{mathrsfs}
\usepackage{pgf,tikz}
\usetikzlibrary{arrows}
\usepackage{mathptmx}
\usepackage{mathrsfs}
\usepackage{epstopdf}



\def\sgn{\mbox{sgn}}

\newtheorem{lemma}{Lemma}[section]
\newtheorem{theorem}{Theorem}[section]
\newtheorem{remark}{Remark}[section]

\newtheorem{definition}{Definition}[section]
\newtheorem{problem}{Problem}[section]
\newtheorem{example}{Example}[section]

\def\bbmat{\begin{bmatrix}}
	\def\ebmat{\end{bmatrix}}

\def\bal{\begin{aligned}}
	\def\eal{\end{aligned}}
\def\bc{\begin{cases}}
	
	\def\ec{\end{cases}}
\def\bi{\begin{itemize}}
	\def\ei{\end{itemize}}
\def\bn{\begin{enumerate}}
	\def\en{\end{enumerate}}
\def\bq{\begin{eqnarray}}
\def\eq{\end{eqnarray}}
\def\bqn{\begin{eqnarray}}
\def\eqn{\end{eqnarray}}
\def\bqta{\begin{eqtarraya}}
	\def\eqta{\end{eqtarraya}}
\def\bqtb{\begin{eqtarrayb}}
	\def\eqtb{\end{eqtarrayb}}
\def\bqtc{\begin{eqtarrayc}}
	\def\eqtc{\end{eqtarrayc}}
\def\be{\begin{equation}}
\def\bes{\begin{equation*}}
\def\ee{\end{equation}}
\def\ees{\end{equation*}}
\def\nn{\nonumber}

\def\zero{{\bf  0}}

\def\bea{\begin{eqnarray}}
\def\eea{\end{eqnarray}}
\def\beann{\begin{eqnarray*}}
\def\eeann{\end{eqnarray*}}

\def\ns{\hspace{-1mm}}

\def\proof{\noindent{\bf{\em Proof:}\ \ }}
\def\QED{\mbox{\rule[0pt]{1.5ex}{1.5ex}}}
\def\endproof{\hspace*{\fill}~\QED\par\endtrivlist\unskip}

\newcommand{\real}{{\mathbb{R}}}

\def\gA{{\cal A}}

\def\gD{{\cal D}}

\def\gE{{\cal E}}

\def\gG{{\cal G}}

\def\gL{{\cal L}}

\def\gN{{\cal N}}

\def\gV{{\cal V}}

\def\bmat{\left[ \begin{array}}
\def\emat{\end{array} \right]}
\def\col{\mbox{col}}

\def\ra{\rightarrow}

\newcommand{\blkd }{\operatorname{blkdiag}}

\def\bsmat{\left[ \begin{smallmatrix}}
\def\esmat{\end{smallmatrix} \right]}

\def\Nous{{\textbf{NOUS}} }
\def\place{{\textbf{place}} }

\definecolor{light-gray}{gray}{0.97}

\begin{document}

\title{Nonovershooting Cooperative Output Regulation for  Linear  Multi-Agent  Systems}
\author{ Robert Schmid  and  Hassan Dehghani Aghbolagh%
\thanks{	Robert Schmid is with the Department of Electrical and Electronic Engineering, University of Melbourne, Parkville, Australia. E-mail: rschmid@unimelb.edu.au. Hassan Dehghani Aghbolagh is with the School of Electrical Engineering and Computing, University of Newcastle, Callaghan, Australia. E-mail: hassan.dehghaniaghbolagh@uon.edu.au} }

\thispagestyle{empty}
\maketitle
\thispagestyle{empty}
\begin{abstract}
		 We consider the  problem of cooperative output regulation for linear multi-agent  systems. A distributed dynamic  output feedback design method is presented that solves the  cooperative  output regulation problem and also ensures that all  agents track  the desired reference signal without overshoot  in their transient response.

\end{abstract}
%
%
\section{Introduction}
In  this  paper, we consider  a family of $N$  linear multi-variable systems ruled by the equations
\be\label{eqsys}
\Sigma_i:\bc
\dot{x}_i(t) =  A_ix_i(t)  +  B_iu_i(t) + E_i w(t),  \; \;\;x_i(0)=x_{i,0} \\
y_i(t)  = C_{y,_i}x_i(t)  +  {D_{y,i}u_i(t)},  +  H_{y,i}w(t)  \\
e_i(t)  =  C_{e,i}x_i(t)  +  D_{e,i}u_i(t)  +  H_{e,i}w(t)

\ec
\ee
where, for all $t \ge 0$, the signal $x_i(t)\in \real^{n_i}$ is the state,
$u_i(t)\in \real^{m_i}$ is the control input,
$y_i(t)\in \real^{p_i}$ is the measured output, and
$e_i(t)\in \real^{\rho_i}$ is  the regulated output of the  $i$-th system,  for $i \in \{1, \dots, N\}$.
The exogenous signal $w(t)\in \real^{q}$ represents a reference signal to be tracked or a disturbance signal to  be rejected, and is assumed to be generated by an exosystem
\be  \label{exos}
\dot w =  Sw,   \quad  w(0) = w_0
\ee
All  matrices appearing in \eqref{eqsys} are appropriate dimensional constant matrices. We assume  the  $N$ agents are divided into two  groups. The first {\it  informed group}  consists of systems  $\Sigma_i$,  for  $i \in \{1,\dots,l\}$,
 that can  access  information about   $w$ from the  measured  output  $y_i$,  which implies  $H_{y,i} \neq  0$. The second {\it uninformed group}   of systems  $\Sigma_i$, for $i \in \{l+1, \dots, N\}$,  for which $H_{y,i} = 0$,  cannot directly access information  about $w$.

The problem of  cooperative output regulation for multi-agent systems involves  designing   control inputs $u_i$ such that the  overall system is  asymptotically stable for the case  $w = 0$, and such that  the tracking errors $e_i$ all converge to  zero, ensuring the outputs of all the agents  converge asymptotically to the desired reference signal.  For the special  case of a  single system ($N=1$), with  access to measurements of the exogenous signal, the  problem reduces to the classic problem of  output feedback regulation. This problem is central to modern control theory.  Solvability conditions and  extensive compilations of results are given in  \cite{Saberi-SS-00}. It is assumed that the measured  output $y_i$  is available for controller design.

The problem of output  regulation of multi-agent systems  has been the subject of a number of papers recently \cite{HGCH}-\cite{SH12b}.  As some of the agents cannot access the  exogenous signal,  the problem cannot be solved by  the  methods of the classical output regulation.
In  \cite{SH12a},  Su  and  Huang  considered  the system (\ref{eqsys}) under the assumption that all   states of  each  system  can  be   measured  and are available for use in  the control   input;  this occurs when  $p_i = n_i$.
They proposed  a  distributed dynamic state feedback  control   scheme and gave conditions  under which  the multi-agent cooperative regulation problem  could be solved.  They showed that their  problem  framework  and controller architecture could accommodate the methods of \cite{HGCH} and \cite{XWL}  as  special  cases.  In \cite{SH12b},  Su  and  Huang extended the  state  feedback methods of \cite{SH12a}  to the case where  $p_i < n_i$  using a distributed  dynamic  measurement feedback control architecture.

For many control  systems there is a need to  avoid undesirable transient phenomena such as high-frequency oscillations   and large magnitudes of the  output  \cite{GL96}.     For a  multi-agent example system, we may consider the  lateral  and directional control of a  research aircraft known  as MuPAL-$\alpha$.  The  flight dynamics of this aircraft were described in \cite{S09},  and   \cite{YW13}  considered the control of  four such  aircraft  within  a  network.   The control objective was for  all the aircraft to simultaneously track a given sideways velocity and a given roll angle.    Exceeding the desired sideways  velocity in  a  platoon may cause some aircraft to  fly too  close together,  and  possibly collide.  If an aircraft exceeds its desired roll angle, its flight may  become unstable and possibly crash.

Thus a desirable transient response should seek  to minimise,  or else  avoid entirely,  overshoot  in the tracking  signal. The  problem of  overshoot is related to  the problem of string stability for automated platoons of vehicles \cite{PWN14,MBL18}.  Such  platoons are usually assumed to be subject to disturbances which should be rejected. Moreover, one of the objectives for them is to track a reference velocity. Obviously, if any of the vehicles in the platoon overshoot in their velocity, collisions might occur.  
Numerous  papers  have  appeared recently seeking to improve the  transient  performance in  the tracking  control of  multi-agent systems, including the  use of consensus  protocols  \cite{PKS17}-\cite{MKD17},  composite nonlinear feedback control \cite{LSY16}, travelling waves \cite{MHS17}, iterative  learning control \cite{YXL16}
 and  transient synchronization \cite{SWA16}.  We note however that none of these  papers  offered a  method for entirely avoiding overshoot in all outputs for  all  the agents.

The design of  control laws to achieve a nonovershooting  step  response for   a single linear time invariant  (LTI)  plant   was  considered  in the  paper  \cite{SN10} by the  first  author of the  present paper.   Several  methods were given  for  the design of a linear state feedback control law   to deliver a nonovershooting step response for an LTI multiple-input multiple-output (MIMO) system.  This requires  the closed-loop  system to  be  stable,  and that the tracking error of the step  response  converges to  zero without changing sign in  any of its components.   In \cite{SN12},  the methods were adapted to the  problem of  avoiding  undershoot in  the step  response,  and  in \cite{SN14}  the methods were used to achieve nonovershooting output regulation.  The  design methods of \cite{SN10} and \cite{SN12} have  been incorporated into a public domain MATLAB$^{\textrm{\tiny{\textregistered}}}$ toolbox, known as \textbf{NOUS} \cite{PS12}.


In  this  paper, we consider how to  combine the  nonovershooting tracking control methods  of  \cite{SN10} with the  distributed control  scheme of \cite{SH12b} to solve the  multi-agent cooperative output regulation  problem in  such a manner that all agents  achieve exact output regulation  with a nonovershooting transient response.  The  principal  contribution of the  paper is to  identify the  necessary system assumptions  and  information required in order for the  control scheme to  deliver a  nonovershooting response.   The authors   believe  that this is the first paper offering  a control scheme to  avoid overshoot  in all outputs of all agents  of  a  multi-agent system. 

  The paper is organised as follows. In Section \ref{secmath},  we introduce some elementary notions from  graph theory that enable us to  define  our  multi-agent problem.  In Section \ref{secpf}, we  introduce the  dynamic measurement output feedback control  architecture  introduced    by  \cite{SH12b},  and  define  our nonovershooting cooperative output regulation problem.  In Section \ref{secnous}, we  briefly discuss  the  nonovershooting controller design methods of  \cite{SN10}. The main  result of the  paper  is  presented in   Section \ref{secps},  where we show how the methods  of \cite{SN10} can  be  employed within the controller architecture of  \cite{SH12b} to  solve  our  problem.

        Section \ref{secex}  demonstrates the  application of the control method  to   the  lateral  and directional control of a network of  research aircraft known  as MuPAL-$\alpha$,  as  discussed in   \cite{YW13}.   Our simulations  demonstrate that the methods  introduced in this  paper can effectively  avoid  overshoot  in all  the outputs  of all the agents  involved in  the  flight simulation. Finally  Section \ref{secconc} offers some concluding thoughts. 

{\bf Notation.} $I_n$ is the $n$-dimensional identity matrix,   and $\zero_{N\times l}$ denotes  an $N \times  l$  matrix with zero  entries.
For a square matrix $A$, we use $\rho(A)$ to denote its spectrum. We say that a square  matrix $A$ is {\it Hurwitz} if $\rho(A)$ lies  within the  open left-hand  complex  plane.
$Re(\lambda)$ denotes the real part of a  complex scalar $\lambda$,  and
$\otimes$  denotes the Kronecker product of matrices.
\section{Mathematical Preliminaries}
\label{secmath}

\subsection{Graph Theory} \label{secgraph}

Graph theory   \cite{graph2} has been widely used to describe the topology of  networked systems by means of vertices and edges. Let $\gG(\gV,\gE, a)$  denote a {\it  weighted digraph}, in which $\gV$  is the finite set of nodes, $\gE$  is the set of directed  edges, and  $a$ represents the set of weights for each edge.   Directed edges have a head  node  and a tail  node.   We use
$(j,i) $ to  denote the edge in $\gE$ directed from tail   node $j$ to head node $i$,  and  $a_{ij}$ denotes the weighting assigned to this  edge.  For  node  $i \in \gV$, we use  $\gN_i$ to denote   all  nodes $j\in \gV$  for which there  exists   an edge  from  tail node  $j$ to head  node $i$.   Thus
\be
\gN_i = \{ j \in \gV:  (j,i)\in \gE \}
\ee
We   refer to the  nodes  in $\gN_i$ as the {\it neighbours}  of node $i$.  A digraph has  a {\it spanning tree}  if there exists at least one node having a directed path to all the other nodes.  The {\it  in-degree} of a node,  denoted by $d_{in}(i) $, is the sum of the  weights of the edges with heads at that node, and is given by
\be
d_{in}(i) =\sum_{j \in \gN_i} \     a_{ij}
\ee
The {\it degree matrix}   of a digraph is a diagonal matrix   $\gD$, whose diagonal  entries are the in-degrees of the  nodes of the digraph from  which it is derived.
The  weighted {\it adjacency matrix}  $\gA$ for a   digraph has entries $\gA_{ij}$  given by
\be \label{eq:adj}
\gA_{ij}=\begin{cases}
	a_{ij}, & (j,i)\in \gE \\
	0,      & \mbox{ otherwise}
\end{cases}
\ee
The information  contained within the degree  and adjacency matrices  of  a  graph  may also be captured within  a single  matrix known  as the {\it  Laplacian matrix},  which is defined as
\be \label{eq:laplace}
\gL=\gD-\gA
\ee

The  $N$ systems of \eqref{eqsys} with the exosystem \eqref{exos} can be viewed as a leader-follower multi-agent system of $N+1$ agents with the exosystem as its leader.   To  model such systems with graphs, we consider a  digraph ${\gG}$   with nodes $\gV = \{0,1,\dots,N\}$  in  which node $0$ represents the  exosystem and the remaining nodes represent the  $N$ agents. The  set of edges  $\gE$ represents the   information  available to the $i$-th   agent for the design of its control law $u_i$.   Thus if $(0,2) \in \gE$, then agent  2 is able to see the state $w$ of  the exosystem,  and  $a_{20} =1$.  If  $(3,2) \notin \gE$, then agent 2  is not able to see the state $x_3$ of  agent 3,  and  $a_{23}=0$.
\begin{lemma}   \cite{SH12b}  \label{SHlem1}
	Let $\gG$ be a  digraph  with Laplacian $\gL$,  and   partition $\gL$ according to
	\be
	\gL = \bmat{c|cc} \zero_{1\times 1} &  \zero_{1\times l} &   \zero_{1\times (N-l)} \\
	\hline
	\gL_{21}  &  \gL_{22}  &  \gL_{23}  \\
	\gL_{31}  &  \gL_{32}  &  \gL_{33}
	\emat
	\ee
	where $\gL_{22} \in \real^{l \times l}$  and  $\gL_{33} \in \real^{(N-l) \times (N-l)}$. Then $\gL_{33}$ is  nonsingular  if and  only if  $\gG$ contains a directed spanning tree  with node 0 as the root. If $\gL_{33}$ is nonsingular, then all its eigenvalues have positive real parts.
\end{lemma}

\subsection{Exponentially decaying sinusoids. }
Our analysis   will require some discussion of the  properties of   exponentially decaying sinusoids.

\begin{definition}  For any positive  integer $n $, let  $\{\mu_i :  i \in \{1, \dots,  n\}\}$, $\{\omega_i: i \in \{1, \dots, n\}\}$,  $\{\alpha_i:  i \in \{1,\dots, n\}\}$ and $\{\beta_i: i \in \{1, \dots,  n\}\}$ be sets of real numbers  such  that   for all $i \in \{1,\dots, n\}$  we  have  $\mu_i < 0$   and  $\omega_i \geq 0$.   Let $f: \real \rightarrow \real$ be given  by
	\be  \label{EDS}
	f(t) = \sum_{i=1}^n \ e^{\mu_i t} [\alpha_i \sin(\omega_i t) + \beta_i \cos(\omega_i t)  ]
	\ee
	Also   let  $\mu <0 $ be given  by
	\be
	\mu = \max\{\mu_i:  i \in \{1, \dots, n\}\}
	\ee
	We say that the scalar function  $f$ is  the  sum of  exponentially decaying  sinusoidal  (SEDS)  functions with  rate  $\mu$. If $v: \real \rightarrow \real^m $ is  a vector-valued function  with $v(t) = [v_1(t)  \dots v_m(t)]^T$,   and  each component   $v_j $ is a SEDS  function of  rate $\mu_j <0$, then we say that  $v$ is  a  SEDS  function with   rate  $\mu =  \max\{\mu_j:  j \in \{1, \dots,  m\}\}$.
	If $f$ is such that  $\omega_i = 0$ for all $ i \in \{1, \dots, n\}$,  then  we say that  $f$ is  the sum of exponentially decaying (SED) functions.
\end{definition}
We  note some straightforward  properties  of SEDS functions; proofs are given in the Appendix.
\begin{lemma}\label{lem61}
	Let $f_1: \real \rightarrow \real$ and  $f_2: \real \rightarrow \real$ be SEDS  functions rate $\mu_1<0$  and  $\mu_2 <0$ respectively. Then $f_1+f_2$  and  $f_1f_2$ are  SEDS  functions with rates   $\mu = \max\{\mu_1,\mu_2\}$,  and  $\mu = \mu_1 + \mu_2 $, respectively.
\end{lemma}

\begin{lemma} \label{lem62}
Consider the linear system
	\bea
	\label{nomsys}
	\begin{array}{rcl}
		\dot{x}(t) \ns&\ns = \ns&\ns A x(t)   +  B u(t) , \quad   x(0)=  x_{0} \\
		y(t) \ns&\ns = \ns&\ns C x(t) + D u(t)
	\end{array}
	\eea
	where $A$ is  Hurwitz. Let  $\lambda_0 = \max\{Re(\lambda): \lambda \in \rho(A)\}$.
	
	\bn
	\item For any $x_0$,   the zero input solution  $x$ and zero input response  $y$  arising from  the input $u$ with   $u(t) = 0$   for all $t \geq 0$  are  SEDS functions with  rate $\lambda_0$.
	\item  If the input  $u$ is a  SEDS function with  rate    $\mu$,  then the zero state  response  $y$ arising from   $x_0=0$ is  a SEDS  function with rate   $\mu$.
	\en
\end{lemma}

\begin{lemma} \label{lem63}
	Let  $f:\real \rightarrow \real$ be  a SEDS  function of the form (\ref{EDS}) with  rate $\mu$,  and  for  some  positive  integer $m$, let $g:\real \rightarrow \real $ be  a SED function given  by
	\be
	g(t) = \sum_{i=1}^m \ \beta_i e^{\lambda_i t}
	\ee
	where   $\{\lambda_1, \dots, \lambda_m\}$  are  distinct negative real numbers satisfying   $\mu_i < \lambda_j$   for all $j \in \{1,\dots, m\}$,  and  $\{\beta_1, \dots, \beta_m\}$ are arbitrary  real numbers.
	Assume $g(t) \neq 0$ for all $ t\geq 0$. Then there exists a positive real number $\delta$ such that $ g(t) + \delta f(t) \neq 0$ for all $ t \geq 0$.
\end{lemma}

\label{secintro}

\section{Problem formulation}
\label{secpf}
Su  and  Huang  in \cite{SH12b}  stated their {\it linear cooperative output regulation problem}  as
\begin{problem} \label{P1}
	For the system (\ref{eqsys})-(\ref{exos})  with digraph $ \gG$,  find suitable control laws $u_i$ of the form (\ref{ulaw1})-(\ref{ulaw3}) for each agent such  that
	\bn
	\item The system matrix of the overall closed loop system is Hurwitz;
	\item For any initial condition $x_{i,0}$,  $\xi_{i,0}$, $\eta_{i,0}$ with $i  \in \{ 1, \dots, N\}$ and  $w_0$,  the  regulated output of the  $i$-th  agent  achieves
	\be
	\lim_{t \rightarrow \infty} \  e_i(t)  =  0,  \quad   i \in \{1, \dots, N\}
	\ee
	\en
\end{problem}
In  this  paper, we consider an extension of this problem, and  seek control laws to  achieve output regulation without  overshoot  in all components of the tracking  error,  for all  agents.
 Since  overshoot  occurs when  the  regulated output changes sign, we  use $e_{i,j}(t)$ to denote the  $j$-th  regulated output component  of the $i$-th agent and define our {\it linear cooperative  nonovershooting output regulation problem}  as follows
\begin{problem}\label{P2}
	For the system (\ref{eqsys})-(\ref{exos}) with digraph $\gG$ and  initial conditions $x_i(0)$ and  $w(0)$,  find suitable linear control laws $u_i$  for each agent that solve Problem \ref{P1} and  also ensure that
$e_{i}(t) \ra 0$ without changing sign in any component, i.e., $\sgn(e_{i,j}(t))$ is constant for all $t \geq 0$, for   every    $j \in \{1, \dots, \rho_i\}$  and  for every   $i \in \{1, \dots, N\}$.
\end{problem}
Next we  discuss the   distributed controller  given in \cite{SH12b}  to solve Problem \ref{P1},  and then  we  review the nonovershooting tracking control methods of \cite{SN10} that we will use to extend the controller of  \cite{SH12b}  to  additionally solve  Problem \ref{P2}.

        \subsection{Distributed dynamic measurement output feedback  control }
Su and Huang \cite{SH12b}   noted   the  following assumptions for each system $\Sigma_i$ in (\ref{eqsys})-(\ref{exos}):
	\bi
	\item[(A.1)]  \label{A1} The matrix $S$ has no  eigenvalues with negative real parts.
	\item[(A.2)]   The pair $(A_i, B_i)$ are  stabilizable,  for  all  $i \in \{1, \dots, N\}$.
	\item[(A.3)]  For every   $i \in \{1, \dots, N\}$, there exist matrices $\Gamma_i$ and $\Pi_i$ satisfying
	\bea
	\Pi_i\,S \ns&\ns = \ns&\ns A_i\,\Pi_i+B_i\,\Gamma_i+E_i \label{Pi1}\\
	0 \ns&\ns = \ns&\ns C_{e,i}\,\Pi_i+D_{e,i}\,\Gamma_i+H_{e,i} \label{CPi2}
	\eea
	\item[(A.4)]  The pairs $\Big( \bmat{cc} C_{y,i} & H_{y,i} \emat, \bmat{cc} A_i & E_i \\ 0 & S \emat \Big)$  are  detectable,  for every
	$i \in \{1, \dots, l\}$.
	\item[(A.5)]  The pairs $(C_{y,i}, A_i) $  are  detectable,  for every  $i \in \{l+1, \dots,N\}$.
	\item[(A.6)] \label{A6} The digraph $ \gG$ contains a directed spanning tree  with node $0$ as its root.
	\ei
	\begin{remark}
Assumptions (A.1)-(A.4)  are standard in the  output regulation literature \cite{Saberi-SS-00},  and are sufficient for the existence of a measurement feedback  controller that can  detect both the  plant state $x_i$ and the  exosystem  state  $w$, for   the  informed agents
  $i \in \{1, \dots, l\}$.   For the  uninformed  agents $i \in \{l+1, \dots, N\}$, (A.5) means that  the plant state $x_i$  is detectable from the measurement output $ y_{i}$, but the exogenous signal $w$  is not  detectable from $y_{i}$ because  $H_{y,i} = 0$.   Hence Problem \ref{P1} cannot be solved by a  decentralized measurement feedback control law. \end{remark}

Using Assumptions (A.1)-(A.5), \cite{SH12b} proposed a distributed dynamic measurement output feedback  controller of  the form:
\be
u_i(t) = F_i\,\xi_i(t)+ G_i\eta_i(t), \quad   i \in \{1, \dots, N\}  \label{ulaw1}
\ee
\be
\label{ulaw2}
\bc
\mbox{if } i  \in \{1, \dots, l\}  \text{ and }  \xi_i(0) = x_{i,0} \\
\bbmat \dot{\xi}_{i}(t) \\ \dot{\eta}_{i}(t) \ebmat =\ns \bbmat A_i & E_i \\ 0 & S \ebmat\bbmat \xi_{i}(t) \\ \eta_{i}(t) \ebmat+\ns\bbmat B_i \\ 0 \ebmat u_i(t)\\ +\bbmat L_{1,i}  \\ L_{2,i} \ebmat \ns\Big(  C_{y,i}  \xi_{i}(t) + D_{y,i}u_i(t) +  H_{y,i}   \eta_{i}(t) -   y_i(t) \Big)
\ec
\ee
\be
\label{ulaw3}
\bc
\mbox{if }  \in       \{l+1, \dots, N\} \text{ and } \eta_i(0) = \eta_{i,0} \\
\bbmat\dot{\xi}_{i}(t) \\ \dot{\eta}_{i}(t) \ebmat = \bbmat A_i & E_i \\ 0 & S \ebmat\,\bbmat \xi_{i}(t) \\ \eta_{i}(t) \ebmat+\bbmat B_i \\ 0 \ebmat \,u_i(t)
\\+\bbmat L_i(C_{y,i} \xi_{i}(t) + D_y u_i(t)  -  y_i(t)) \\  \gamma\sum_{j=1}^N  \  a_{ij} (\eta_{j}(t) - \eta_{i}(t) ) \ebmat,  \quad
\ec
\ee
where  $\gamma >0$,  $F_i \in \real^{m_i \times n_i}$,  $G_i \in \real^{m_i \times q}$, $L_{1,i} \in \real^{n_i \times p_i}$, $L_{2,i} \in \real^{q \times p_i}$  and  $L_i \in \real^{n_i \times p_i}$ are gain matrices,  and  the  parameters  $a_{ij}$ are the entries of the adjacency  matrix of $\gG$.

Thus   the  control  law (\ref{ulaw3}) combines a  distributed  observer with a  Luenberger observer, and  \cite{SH12b} described  the  controller  (\ref{ulaw1})-(\ref{ulaw3}) as a  {\it distributed dynamic measurement output feedback  controller}.
Their main  result was to show that  their controller can solve   Problem \ref{P1}:
\begin{theorem}[\cite{SH12b}, Theorem  1]
	Under Assumptions (A.1)-(A.5), the  cooperative output regulation Problem \ref{P1} is solvable by a distributed dynamic measurement output feedback  control law of the form (\ref{ulaw1})-(\ref{ulaw3}),  if and  only if Assumption (A.6) holds.
\end{theorem}

\subsection{Nonovershooting tracking controller design methods}
\label{secnous}

Schmid and  Ntogramatzidis \cite{SN10} used  state feedback  control  design methods  to deliver  a nonovershooting step response for a single LTI plant ($N=1$).  Here we discuss how these may be applied to multi-agent system  $\Sigma_i$.  We consider the {\em nominal  systems} that  arise when the  exosystem  (\ref{exos})  is excluded from consideration ($S=0$ and $w(0)=0$).  In this case each  agent  in (\ref{eqsys}) simplifies to
\be\label{nomsysi}
\Sigma_{i,nom}:
\bc
\dot{\tilde x}_i(t)  = A_i\,\tilde x(t) + B_i\,\tilde u(t), & \tilde x_i(0)=\tilde x_{i,0}\\
\tilde y_i(t)  =  C_{y,_i}\,\tilde x_i(t)  +  {D_{y,i}\,\tilde u_i(t)} & \\
\tilde e_i(t)  =  C_{e,i}\,\tilde x_i(t)  +  D_{e,i}\,\tilde u_i(t), &      i \in \{ 1, \dots, N\}
\ec
\ee

 \cite{SN10} gave several methods for the design of a linear state feedback control law $\tilde u = F \tilde x$  to deliver a nonovershooting step response for a system in the form (\ref{nomsysi}). This requires ensuring   that the closed-loop  system is asymptotically stable,  and the  tracking error $\tilde e_i$ converges to  zero without overshoot;  this  implies  $\tilde e_{i,j}(t) \ra 0$ as  $t \ra \infty$ without changing sign  in all output components  $j \in \{1, \dots, \rho_i\}$.

The design method  assumed that  initial condition  $\tilde x_{i,0} \neq  0$ of each nominal system (\ref{nomsysi}) is  known and available for  use in the controller design.  The closed-loop eigenvalues to  be assigned by the state feedback   are to be selected from within a user-specified interval of the negative real line. The algorithm selects candidate sets  of distinct closed-loop eigenvalues from within the specified interval and then associates them with candidate sets of closed-loop eigenvectors in such a way that only a small number (generally one or two, or at most three) of the closed-loop modes contribute to each output component.  The  candidate  eigenvalues are associated with  candidate eigenvectors and eigendirections by solving a system of equations involving the   Rosenbrock matrix of the   system (\ref{nomsysi}).  These eigenvectors and  eigendirections are used to obtain a  feedback matrix via  Moore's pole placement algorithm \cite{M}.

The  error signal  $\tilde e(t) $ is then formulated in terms of the candidate set of eigenvectors and a test is used to determine if the system response is nonovershooting in all components. If the test is not successful, then a new candidate set  of eigenvalues  within  the specified interval is  chosen, and the process is repeated. The tests are analytic in nature, and do not require simulating the system response to test for overshoot.

The nonovershooting controller design method can be applied to multiple-input  multiple-output systems,  and  these may be of   non-minimum phase.  The designer has considerable freedom  to  select the desired closed-loop eigenvalues, in order to accommodate requirements on the  convergence rate,  or to  avoid actuator saturation.  The algorithm involves a search for suitable feedback matrices to  deliver a nonovershooting response,  and a successful  search cannot be guaranteed for  any given  system,  for any  given initial condition.   \cite{SN10} gives some discussion of the  circumstances in  which a successful search is likely.     The condition was  that
\be   \label{zerocond}  n -3p \geq z  \ee
where $n$  is the number of states,  $p$ is the number of inputs/outputs,  and $z$ is the  number of minimum-phase zeros. 


In this paper,  we shall assume  the existence of  feedback matrices that yield a nonovershooting response for the  nominal  system of  each agent $\Sigma_{i,nom} $ with initial condition $\tilde x_{i,0}$ in (\ref{nomsysi}):
	\bn
	\item[(A.7)]    A feedback  gain matrix $F_i$
	exists such that   the  eigenvalues of  $A_i+B_i F_i$  are real,  distinct  and negative,  and
	\item[(A.8)]  applying  the control law  $\tilde u_i = F_i \tilde x$ to  $\Sigma_{i,nom}$,  with initial  condition $\tilde x_{i,0} = x_{i,0} -\Pi_i w_0 $, yields nonovershooting  regulated outputs $\tilde e_i$.
	\en
 We  note that condition (A.8) might be difficult to satisfy for some  multi-agent systems  from  some initial  conditions, because it seeks to  avoid  overshoot  in  all the  output components  of all  agents.  In many  practical  problems it may not be essential to avoid overshoot in all outputs,  and  in  such  cases  it becomes easier to  find suitable  feedback matrices to  deliver a   nonovershooting  response for the outputs where avoiding overshoot  is  important. The  methods  of  \cite{SN10} can  accommodate  nonovershooting  requirements for only a selection of the outputs, and the   \textbf{NOUS}   toolbox \cite{PS12} offers  an option  for the  user to specify whether or not  overshoot  is to  be avoided for  each  output component.

\section{Problem  Solution} \label{secps}

Here  we present the main results of our paper, providing a  solution for  Problem \ref{P2}  under Assumptions  (A.1)-(A.8).   Thus we  assume  we have,  for  any initial  condition $\tilde x_{i}(0)$ and $w(0)$,   gain matrices $F_i$  such  that  applying the   control law $\tilde  u_i = F_i \tilde  x_i$ to the  nominal  system  $\Sigma_{i,nom} $ of each   agent  yields a  nonovershooting  response,  from the initial  condition  $\tilde  x_{i,0} =    x_{i,0} - \Pi_i w_0$.  Our task is  to obtain suitable gain matrices  $G_i $, $L_{1,i}$, $L_{2,i}$,   and  $L_i $ and  parameter  $\gamma$  so  that the control laws  (\ref{ulaw1})-(\ref{ulaw3})  will solve Problem \ref{P2}. Firstly we introduce
\be \label{Gi}
G_i = \Gamma_i - F_i \Pi_i, \quad \mbox{for  $i \in \{1,  \dots,  N\}$ }
\ee
Define $\lambda_0 = \min\{\lambda: \lambda \in \rho(A_i+B_iF_i)   $ for any $ i \in \{1,\dots, N\}$;  then $\lambda_0$ provides a  lower bound on eigenvalues of all  the closed-loop state matrices $ A_i + B_i F_i$.  Next  we   chose  $\mu_0  < \lambda_0$ and  obtain  suitable observer  gains $L_{1,i}$,  $L_{2,i}$ for  $i \in \{1,\dots, l\}$,   and  $L_i$ for $i \in \{l+1,\dots, N\}$, such  that the  matrices
\be \label{Acci}
A_{cc,i} =  \bbmat A_i + L_{1,i}C_{y,i} & E_i + L_{1,i} H_{y,i} \\
L_{2,i} C_{y,i}  & S + L_{2,i} H_{y,i} \ebmat
\mbox{ and }      A_i+ L_i C_{y,i}
\ee have distinct  stable eigenvalues all lying to the left of  $ \mu_0$, i.e. for all  $\mu \in \rho(A_{cc,i})$,
and   for all  $\mu \in \rho(A_i+ L_i C_{y,i} )$, we have $Re(\mu) \leq \mu_0$. Thus  $\mu_0$ provides an upper bound  on the real part of the eigenvalues  of all the closed-loop observer matrices.
By Lemma \ref{SHlem1} and  (A.6), we  know that the  real parts of   the eigenvalues  of $\mathcal L_{33}$ are  positive, so  there exists  $\gamma >0$ such that
\be\label{gammadef}
\bal
\max \big\{Re(&\lambda_i(S)  - \gamma\lambda_j(\gL_{33})): \\& i \in \{1, \dots, q\}, j \in \{1,\dots, N-l\} \big\}  \leq \mu_0
\eal
\ee
where $\lambda_i(S)$ and  $\lambda_j(\gL_{33})$ denote the eigenvalues of $S$ and $\gL_{33}$ respectively.

Next we introduce some notation  that will allow us to  compactly  represent the  overall  closed-loop system of (\ref{eqsys})-(\ref{exos})  under  control laws (\ref{ulaw1})-(\ref{ulaw3}).   For  $i \in \{1, \dots, l\}$, we   define
$
 \bar A   =    \blkd (A_1,   \dots,   A_l), \
 \bar B   =    \blkd (B_1,   \dots,   B_l), \
 \bar C_y   =    \blkd (C_{y,  1},    \dots,  C_{y,  l}), \
 \bar C_e   =    \blkd (C_{e,  1}, \   \dots, \  C_{e,  l}), \
 \bar D_y   =    \blkd (D_{y,  1}, \   \dots, \  D_{y,  l}), \
    \bar D_e   =    \blkd (D_{e,  1}, \   \dots, \  D_{e,  l}), \
    \bar E   =    \blkd (E_1, \   \dots, \  E_l), \
 \bar H_y   =    \blkd (H_{y,  1}, \   \dots, \  H_{y,  l}), \
 \bar H_e   =    \blkd (H_{e,  1}, \   \dots, \  H_{e,  l}), \
 \bar F   =    \blkd (F_{1},    \dots, \  F_{l}), \
  \bar G   =    \blkd (G_{1}, \   \dots, \  G_{l}), \
 \bar L_1   =    \blkd (L_{1,  1}, \   \dots, \  L_{1, l}), \
 \bar L_2   =    \blkd (L_{2,  1}, \   \dots, \  L_{2,  l}), \
 \bar S    =    I_l  \otimes S, \
  \bar  \Pi  =   \blkd ( \Pi_1, \   \dots, \   \Pi_l), \
 \bar  \Gamma  =   \blkd ( \Gamma_1,   \dots,    \Gamma_l), \
 \bar x    =      \col(x_1, \   \dots,   x_l), \
 \bar e    =     \col(e_1, \   \dots,   e_l), \
 \bar w  =   \col(w, \   \dots, \   w), \
\bar \xi     =     \col(\xi_{1} \dots,   \xi_{l}), \
\bar \eta     =    \col(\eta_{1} \dots,   \eta_{l}) , \
\bar \varepsilon_{1}    =   \bar \xi  - \bar x, \
\bar \varepsilon_{2}     =     \bar \eta  -  w, \
\bar \varepsilon    =   [\varepsilon_1, \   \varepsilon_2]^T$.

For $i \in \{l+1, \dots, N\}$, we similarly  define matrices
$\hat A$, $\hat B$, $\hat C_y$, $\hat C_e$, $\hat D_y$, $\hat D_e$, $\hat E $,  $\hat H_e$, $\hat G$,
$\hat L = \blkd (L_{l+1}, \dots, L_{N}) $,  $\hat S = I_{N-l} \otimes S$, $\hat \Pi$, $\bar \Gamma$,  and  variables $\hat x$, $\hat e$, $\hat w$, $\hat \xi$,  $\hat \eta$,
$\hat  \varepsilon_1$,   $\hat  \varepsilon_2$  and  $\hat \varepsilon$.    In this form,  the regulator equations  (\ref{Pi1})-(\ref{CPi2}) become
\bq
\bar \Pi\bar S \ns&\ns = \ns&\ns \bar  A\,\bar \Pi+\bar B\,\bar \Gamma+\bar E \label{barPi1}\\
0 \ns&\ns = \ns&\ns \bar C_e\,\bar \Pi+\bar D_{e}\,\bar \Gamma+\bar H_{e} \label{barCPi2}  \\
\hat \Pi\hat S \ns&\ns = \ns&\ns \hat  A\,\hat \Pi+\hat B\,\hat \Gamma+\hat E \label{hatPi1}\\
0 \ns&\ns = \ns&\ns \hat C_e\,\hat \Pi+\hat D_{e}\,\hat \Gamma+\hat H_{e} \label{hatCPi2}
\eq
\begin{theorem} \label{mainthm}
	Consider the  multi-agent cooperative system $\Sigma_i$ in (\ref{eqsys}) under  assumptions (A.1)-(A.6) and  initial conditions $x_{i,0}$ and  $w_0$. Assume that a distributed  dynamic measurement output feedback  controller of the form  (\ref{ulaw1})-(\ref{ulaw3}) has been obtained that satisfies (A.7)-(A.8)   and (\ref{Gi})-(\ref{gammadef}) for all $i \in \{1,\dots, N\}$.
	Then this  control law solves  Problem \ref{P2},  provided the  initial  estimator  error  $(\bar \varepsilon(0), \hat \varepsilon(0))$ is sufficiently small.
\end{theorem}
\proof  Firstly we obtain expressions for the closed loop system under the  controller (\ref{ulaw1})-(\ref{ulaw3}). For  the informed agents  $i \in \{1,\dots, l\}$,  the tracking error  dynamics are given  by
\be
\bal
\dot{\bar \varepsilon}(t)=&   \bmat{cc} \bar A & \bar  E \\ 0 & \bar  S \emat\,\bmat{c} \bar \xi(t) \\ \bar \eta(t) \emat+\bmat{c} \bar B \\ 0 \emat \, \bar u(t) \\& +\bbmat \bar L_{1}\\  \bar L_{2} \ebmat \Big(  \bar C_{y} \bar \xi(t) + \bar D_{y}\bar u(t) + \bar H_{y}  \bar \eta(t) -  \bar  y(t) \Big)  \\&
-   \bmat{cc} \bar A & \bar E \\ 0 & \bar S \emat\,\bmat{c} \bar x(t) \\ \bar \eta(t) \emat-\bmat{c}\bar B \\ 0 \emat \,\bar u(t)  \\
= &    \bmat{cc} \bar A & \bar  E \\ 0 & \bar  S \emat\,\bmat{c} \bar \epsilon_1(t) \\ \bar \epsilon_2(t) \emat
\\&+  \bbmat \bar L_{1}  \\ \bar L_{2} \ebmat \Big(  \bar C_{y} \bar \xi(t) + \bar D_{y}\bar u(t)   + \bar H_{y}  \bar \eta(t) \\&\qquad\quad\;- ( \bar   C_{y}\bar x(t) + \bar D_{y}\bar u(t)  + \bar  H_{y}\bar w(t) )\Big)   \\
= & \bmat{cc} \bar A+\bar L_1\bar C_y & \bar E +\bar L_1\bar H_{y} \\  \bar L_2\bar C_y & \bar S+\bar L_2 \bar H_{y} \emat\,\bar \varepsilon(t)  \nn
\eal
\ee
The state and exosystem  dynamics are given by
\be
\bal
\bmat{c}  \dot{\bar x}(t) \\  \dot{\bar w}(t) \emat =& \bmat{cc} \bar A & \bar E\\ 0 &\bar S \emat\,\bmat{c}\bar x(t) \\ \bar w(t) \emat+\bmat{c} \bar  B \\ 0 \emat \bar u(t) \\
=& \bmat{cc} \bar A & \bar E \\ 0 & \bar S \emat\,\bmat{c}\bar  x(t) \\ \bar w(t) \emat+\bmat{c} \bar B \\ 0 \emat \,(\bar F\,\bar \xi(t)+\bar G\,\bar \eta(t)) \\
=& \bbmat \bar A+\bar B\bar F \ns&\ns \bar E +\bar B\bar G \\ 0 \ns&\ns \bar S \ebmat\,\bbmat \bar x(t) \\ \bar w(t) \ebmat+\bbmat \bar B\bar F & \bar B\bar G \\ 0 & 0 \ebmat \,\bbmat \varepsilon_1(t) \\ \varepsilon_2(t) \ebmat \nn
\eal
\ee
and hence the  closed-loop system for   agents  $i \in \{1, \dots, l\}$ is
\beann
\bbmat
\dot{\bar x}(t) \\ \dot{\bar w}(t) \\  \dot{\bar \varepsilon}_1(t) \\ \dot{\bar \varepsilon}_2(t) \ebmat  =
\bmat{cc|cc}
\bar A+\bar B\bar F \ns&\ns \bar E +\bar B\bar G \ns&\ns \bar B\bar F \ns&\ns \bar B\bar G \ns\\
0 \ns&\ns\bar S \ns&\ns 0 \ns&\ns 0 \ns\\
\hline
0 \ns&\ns 0 \ns&\ns \bar A+\bar L_1 \bar C_y \ns&\ns\bar  E +\bar L_1\bar H_{y}\ns\\
0 \ns&\ns 0 \ns&\ns \bar L_2\bar C_y \ns&\ns \bar S+ \bar L_2\bar H_{y} \ns\emat \ns \ns\bbmat
\bar{x}(t) \\ \bar{w}(t) \\  {\bar \varepsilon}_1(t) \\ {\bar \varepsilon}_2(t) \ebmat \\
\eeann
\beann
\bar e(t)  \ns&\ns = \ns&\ns \bmat{cc|cc}
\bar C_e +\bar D_{e}\bar F & \bar D_{e}\bar G + \bar H_{e} & \bar D_{e}\bar F & \bar D_e\bar G \emat \bbmat
\bar{x}(t) \\ \bar{w}(t) \\  {\bar \varepsilon}_1(t) \\ {\bar \varepsilon}_2(t) \ebmat
\eeann
Introducing coordinates $\bar z(t) = \bar x(t)-\bar \Pi\,\bar w(t) $ and using \eqref{barPi1}, we obtain
\be
\bal
\dot{ \bar z}(t)
=&  \dot{ \bar x}(t) - \bar \Pi \dot{ \bar w}(t)\nn  \\
=&  \dot {\bar x}(t) - \bar \Pi\bar  S \bar w(t)  \nn\\
=&  (\bar A + \bar B\bar F)\bar x(t)+(\bar E +\bar B\bar G -\bar \Pi \bar S)\bar w(t)\nn +\nn \bar B\bar F\bar \varepsilon_1(t) \ns +\bar B\bar G \bar \varepsilon_2(t)\\
=&  (\bar A + \bar B\bar F)\bar x(t) - (\bar A + \bar B\bar F)\bar \Pi\bar w(t)
+ \bar B\bar F\bar \varepsilon_1(t) + \bar B\bar G \varepsilon_2(t),   \\
=& (\bar A + \bar B \bar F)\bar z(t)+ \bar B\bar F\bar \varepsilon_1(t) +\bar B\bar G \bar \varepsilon_2(t)
\eal
\ee
Also
\be
\bal
(\bar C_e &+\bar D_{e}\bar F)\bar x(t) +(\bar D_{e}\bar G + \bar H_{e})\bar w(t)\\
= & (\bar C_e +\bar D_{e}\bar F)\bar x(t) + (\bar D_{e}\bar \Gamma  - \bar D_{e}\bar F\bar \Pi + \bar H_{e})\bar w(t) \nn \\
= & \bar C_e \bar x(t) + \bar D_{e}\bar F (\bar x(t) - \bar \Pi \bar w(t)) +( \bar D_{e}\bar \Gamma +\bar H_{e}) \bar w(t) \nn \\
= & \bar C_e \bar z(t) +  \bar D_{e}\bar F\bar z(t) + (\bar  C_e \bar \Pi + \bar D_{e}\bar \Gamma + \bar H_{e})\bar  w(t) \nn \\
= & (\bar C_e + \bar D_{e}\bar F)\bar z(t)
\eal
\ee
by (\ref{barCPi2}).
Hence we may write the closed loop system as
\bq
\dot{\bar z}(t) &=& (\bar A+\bar B\bar F)\,\bar z(t) + [ \bar B\bar F \ \ \bar B\bar G  ] \bar \varepsilon(t), \;\; \bar z(0)=\bar z_0   \label{clinfa} \\
\dot {\bar \varepsilon}(t) & = & \bar A_{cc} \bar \varepsilon(t), \quad  \bar \epsilon(0)= \bar \epsilon_0  \label{clinfb} \\
\bar e(t) &=& (\bar C_e+\bar D_{e}\,\bar F)\,\bar z(t) + [\bar  D_{e}\bar F \ \ \bar D_e\bar G)]\bar \varepsilon(t)
\label{clinfc}
\eq
where
\be \bar{A}_{cc} =    \bmat{cc} \bar A + \bar{L}_{1}\bar{C}_{y} & \bar{E} + \bar{L}_{1} \bar{H}_{y} \\
\bar{  L}_{2} \bar{C}_{y}  & \bar S +  \bar{L}_{2}  \bar{H}_{y} \emat
\ee
Secondly  we consider  the uninformed agents for  $i \in \{l+1,\dots, N\}$  and denote the estimation error as
\bea
\hat \varepsilon(t) =  \bmat{c} \hat \varepsilon_1(t) \\ \hat \varepsilon_2(t) \emat =
\bmat{c} \hat \xi(t) - \hat  x(t) \\ \hat \eta(t) - \hat w
\emat
\eea
From Lemma  2 in \cite{SH12b}, we  know that
\be
\dot{\hat{\varepsilon}}_2 =  \hat S - \gamma(\gL_{33} \otimes I_q) \hat \epsilon_2 - \gamma(\gL_{32} \otimes I_q) \bar \epsilon_2
\ee
so that
\bes
\bal
\dot{\hat \varepsilon}(t)=&   \bmat{cc} \hat A & \hat  E \\ 0 & \hat  S \emat\,\bmat{c} \hat \xi(t) \\ \hat \eta(t) \emat+\bmat{c} \hat B \\ 0 \emat \, \hat u(t) \\& -
\bmat{c} \hat L(\hat{C}_{y}  \hat{\xi}(t) + \hat{D}_y \hat{u}(t)  - \hat {y}) \\    \gamma(\gL_{33} \otimes I_q) \hat \epsilon_2 + \gamma(\gL_{32} \otimes I_q) \bar \epsilon_2   \emat  \\
& -   \bmat{cc} \hat A & \hat E \\ 0 & \hat S \emat\,\bmat{c} \hat x(t) \\ \hat \eta(t) \emat-\bmat{c}\hat B \\ 0 \emat \,\hat u(t)  \\
= &   \bmat{cc} \hat A & \hat  E \\ 0 & \hat  S \emat\,\bmat{c} \hat \epsilon_1(t) \\ \hat \epsilon_2(t) \emat
\\&- \bmat{c} \hat L ( \hat  C_{y}\hat{\xi}(t) + \hat D_{y}\hat u(t)  - \hat{C}_y \hat{x}(t) - \hat {D}_y \hat{u}(t) ) \\
\gamma(\gL_{33} \otimes I_q) \hat \epsilon_2 + \gamma(\gL_{32} \otimes I_q) \bar \epsilon_2  \emat
\\
= & \bbmat \hat A+\hat L\hat C_y & \hat E  \\ 0 & \hat S  -  \gamma (\gL_{33} \otimes I_q)   \ebmat\,\hat \varepsilon(t)
-\ns \bbmat 0  \\  \gamma(\gL_{32} \otimes I_q) \bar \epsilon_2  \ebmat
\eal
\ees
It  follows that the  closed loop-system for agents $i \in \{l+1, \dots, N\}$ is
\beann
\hspace{-2mm}
\bbmat
\dot{\hat x}(t) \\ \dot{\hat w}(t) \\  \dot{\hat \varepsilon}_1(t) \\ \dot{\hat \varepsilon}_2(t) \ebmat
&\hspace{-6mm} = &
\hspace{-6mm}
\bmat{cc|cc}
\hat A+\hat B\hat F \ns&\ns \hat E +\hat B\hat G \ns&\ns \hat B\hat F \ns&\ns \hat B\hat G \ns\\
0 \ns&\ns \hat S \ns&\ns 0 \ns&\ns 0 \ns\\
\hline
0 \ns&\ns 0 \ns&\ns \hat A+\hat L \hat C_y \ns&\ns\hat  E \ns\\
0 \ns&\ns 0 \ns&\ns 0 \ns&\ns \hat S -  \gamma (\gL_{33} \otimes I_q) \ns\emat
\hspace{-1.5mm}
\bbmat
\hat{x}(t) \\ \hat{w}(t) \\  {\hat \varepsilon}_1(t) \\ {\hat \varepsilon}_2(t) \ebmat    \\
& &- \bbmat  0 \\ 0 \\ 0 \\  \gamma(\gL_{32} \otimes I_q) \bar \epsilon_2  \ebmat    \\
\hat e(t) & \hspace{-4mm}  = &
\hspace{-4mm} \bmat{cc|cc}
\hat C_e +\hat D_{e}\hat F & \hat D_{e}\hat G + \hat H_{e} & \hat D_{e}\hat F & \hat D_e\hat G \emat \bmat{c}
\hat {x}(t) \\ \hat {w}(t) \\  {\hat \varepsilon}_1(t) \\ {\hat \varepsilon}_2(t) \emat
\eeann

Introducing the change of coordinates $\hat  z(t) = \hat  x(t)-\hat \Pi\,\hat  w(t) $, we obtain the closed-loop system in the form
\bq
\dot{\hat z}(t) &=& (\hat A+\hat B\hat F)\,\hat z(t) + [ \hat B\hat F \ \ \hat B\hat G  ] \hat \varepsilon(t), \  \hat z(0)=\hat z_0     \label{clnoninfa} \\
\dot {\hat \varepsilon}(t) & = & \hat A_{cc} \hat \varepsilon(t)  - \bmat{c}    0 \\   \gamma(\gL_{32} \otimes I_q) \emat \bar \epsilon_2 , \  \hat \epsilon(0)= \hat \epsilon_0      \label{clnoninfb} \\
\hat e(t) &=& (\hat C_e+\hat D_{e}\,\hat F)\,\hat z(t) + [\hat  D_{e}\hat F \ \ \hat D_e\hat G)]\hat \varepsilon(t)    \label{clnoninfc}
\eq
where
\be  \label{hatAcc}
\hat{A}_{cc} =    \bmat{cc} \hat A + \hat{L}\hat{C}_{y} & \hat{E}  \\
0     & \hat S  - \gamma(\gL_{33} \otimes I_q) \emat
\ee
Next   we consider the form of the outputs arising from these closed-loop systems. Firstly we consider  (\ref{clinfa})-(\ref{clinfc}) for the  informed agents.   We may  decompose the  state vector $\bar z$  according to   $\bar z =  \bar z_A  + \bar z_B$  where  $\bar z_A $ and $\bar z_B$ are the  zero input solution and zero  state solutions,  respectively.  Similarly we  can  decompose the  output  $\bar e$  into  $\bar e =  \bar e_A  + \bar e_B$,  the   zero input response and zero  state responses,  respectively.  By Assumptions (A.7)-(A.8),   for each agent, $A_i + B_i F_i$ is Hurwitz with real,  negative and distinct eigenvalues,  and the output $\tilde e_{i}$ of the nominal  system $\Sigma_{i,nom}$ in (\ref{nomsysi})  from initial  condition $x_{i,0} - \Pi_i w_0$ is  nonovershooting.  Since  $\bar e_A$ is composed of the   $\tilde e_i$   outputs from all the  informed  agents,  we conclude that $\bar z_A$ and $\bar e_A$ are  SED  functions  and  $\bar e_A(t) \ra  0$  as  $t \ra \infty$ without changing sign in  any component.

Considering   the  error dynamics for $\bar \varepsilon$ in (\ref{clinfb}), we know  by  (\ref{Acci})  that $\bar A_{cc}$  is  Hurwitz  and satisfies $\max\{\mu:  \mu  \in \rho(\bar A_{cc})\}  \leq \mu_0$.  By Lemma \ref{lem62}(i),  $\bar \varepsilon$ is a  SEDS function with rate $\mu_0$, and  $|\bar \varepsilon(t)|   \leq k_1|\bar \varepsilon_0|$,  for some  $k_1 >0$.
As $\bar \varepsilon$ is the input for
(\ref{clinfa}), by Lemma \ref{lem62}(ii),  we conclude that $\bar e_B$ is a  SEDS functions with rate at most $\bar \mu$, and  $|\bar e_B(t)|   \leq k_2k_1|\bar \varepsilon_0|$,  for some  $k_2 >0$.    We may now apply  Lemma \ref{lem63} with $g =  \bar e_A$ and
$f =  \bar e_B$. Provided  $|\bar \varepsilon_0|$ is sufficiently  small, we have  $\bar e(t) \ra  0$ as $t \ra \infty$  without changing sign in  any component.

Next   we consider the form of the outputs arising from these closed-loop systems. Firstly we consider  (\ref{clinfa})-(\ref{clinfc}) for the  informed agents.   We may  decompose the  state vector $\bar z$  according to   $\bar z =  \bar z_A  + \bar z_B$  where  $\bar z_A $ and $\bar z_B$ are the  zero input solution and zero  state solutions,  respectively.  Similarly we  can  decompose the  output  $\bar e$  into  $\bar e =  \bar e_A  + \bar e_B$,  the   zero input response and zero  state responses,  given  by
\bq
\bar e_A(t) &=& (\bar C_e+\bar D_{e}\,\bar F)\,\bar z(t)   \label{bareA} \\
\bar e_B(t) &=&  [\bar  D_{e}\bar F \ \ \bar D_e\bar G)]\bar \varepsilon(t) \label{bareB}
\eq 
 By Assumptions (A.7)-(A.8),   for each agent, $A_i + B_i F_i$ is Hurwitz with real,  negative and distinct eigenvalues,  and the output $\tilde e_{i}$ of the nominal  system $\Sigma_{i,nom}$ in (\ref{nomsysi})  from initial  condition $x_{i,0} - \Pi_i w_0$ is  nonovershooting.  Since  $\bar e_A$ is composed of the   $\tilde e_i$   outputs from all the  informed  agents,  we conclude that $\bar z_A$ and $\bar e_A$ are  SED   functions  and  $\bar e_A(t) \ra  0$  as  $t \ra \infty$ without changing sign in  any component.

Considering   the  error dynamics for $\bar \varepsilon$ in (\ref{clinfb}), we know  by  (\ref{Acci})  that $\bar A_{cc}$  is  Hurwitz  and satisfies $\max\{\mu:  \mu  \in \rho(\bar A_{cc})\}  \leq \mu_0$.  By Lemma \ref{lem62}(i),  $\bar \varepsilon$ is a  SEDS function with rate $\mu_0$, and  $|\bar \varepsilon(t)|   \leq k_1|\bar \varepsilon_0|$,  for some  $k_1 >0$.
As $\bar \varepsilon$ is the input for
(\ref{clinfa}), by Lemma \ref{lem62}(ii),  we conclude that $\bar e_B$ is a  SEDS functions with rate at most $\bar \mu$.
  We may now apply  Lemma \ref{lem63} with $g =  \bar e_A$ and
$f =  \bar e_B$    to  obtain $\delta  >0$ such that
\be
\bar e_A(t)   + \delta \bar e_B(t)  >0
\ee
From (\ref{clinfb}) and  (\ref{bareB}), we see that  $\bar e_B$ is  linearly dependent upon  the  initial condition $\bar \varepsilon_0$,  and  hence for suitably small $|\bar \varepsilon_0|$,   we  have $\bar e_A(t)   +  \bar e_B(t)  >0$,
Thus   $\bar e(t) = \bar e_A(t)   +\bar  e_B(t)  \ra  0$ as $t \ra \infty$  without changing sign in  any component.

Next we consider the  uninformed agents (\ref{clnoninfa})-(\ref{clnoninfc}) for  $i \in \{l+1, \dots, N\}$.  We  again  decompose  the state    vector as  $\hat z =  \hat z_A  + \hat z_B$  where  $\hat z_A $ and $\hat z_B$ are the  zero input and zero  state solutions,  respectively.  Similarly we  have   $\hat e =  \hat e_A  + \hat e_B$ for the   zero input response and zero  state responses,  given  by
\bq
\hat e_A(t) &=& (\hat C_e+\hat D_{e}\,\hat F)\,\hat z(t)       \label{hateA}    \\
\hat e_B(t) &=&  [\hat  D_{e}\hat F \ \ \hat D_e\hat G)]\hat \varepsilon(t)     \label{hateB}
\eq 
 Again by assumptions  (A.7)-(A.8),   we  have that  for each agent, $A_i + B_i F_i$ is Hurwitz with negative, real  and distinct  eigenvalues,  and the output $\tilde e_{i}$  of the  nominal  system $\Sigma_{i,nom}$ from initial  condition $x_{i,0} - \Pi_i w_0$ is  nonovershooting.
Hence   $\hat z_A$ and $\hat e_A$ are  SED   functions  and  $\hat  e_A(t) \ra  0$ as $t \ra \infty$ without changing sign in any component.

Considering the  error dynamics for  $\hat \varepsilon$ in   (\ref{clnoninfb}),  we know  by  (\ref{Acci}) that  $\hat A + \hat L_1 \hat C_y$  is  Hurwitz  and all its eigenvalues satisfy  $Re(\mu) \leq   \mu_0$.  From Lemma  2 of \cite{SH12b}, we have
\be
\bal
\rho(  \hat S  - \gamma(\gL_{33}& \otimes I_q))  =
\{\lambda_i(S) - \gamma \lambda_j(\mathcal{L}_{33}): \\& \qquad i \in \{1,\dots, q\}, \ j \in \{1,\dots, N-l\}\}
\eal
\ee
As $ \gamma $ satisfies (\ref{gammadef}),  we know that  $\hat S  - \gamma(\gL_{33} \otimes I_q)$ is  Hurwitz,  and all its eigenvalues satisfy  $Re(\mu) \leq   \mu_0$.   We conclude that $\hat{A}_{cc} $  in (\ref{hatAcc}) is  Hurwitz   and  its eigenvalues satisfy   $Re(\mu) \leq  \mu_0$.

Decomposing $\hat  \varepsilon  = \hat \varepsilon _A + \hat \varepsilon_B$ into its  zero input  and zero state solutions, we observe from Lemma \ref{lem62}(i) that  $ \hat \varepsilon_A$ is a  SEDS function with  rate at most  $\mu_0$.
From above we know that  $\bar \varepsilon$, and  hence also $\bar \varepsilon_2$, are  SEDS functions  with  rate $\mu_0$.
 Thus by Lemma \ref{lem62}(ii), $\hat \varepsilon_B$ is also  a SEDS function   with  rate $\mu_0$.

As $\hat  \varepsilon$ is the input for
(\ref{clnoninfa}), by Lemma \ref{lem62}(ii),  we conclude that $\hat e_B$ is a  SEDS functions with rate at most $\bar \mu$, 
  We may now apply  Lemma \ref{lem63} with $g =  \hat e_A$ and
$f =  \hat e_B$  to  obtain $\delta  >0$ such that
\be
\hat e_A(t)   + \delta \hat e_B(t)  >0
\ee
From (\ref{clnoninfb}) and  (\ref{hateB}), we see that  $\hat e_B$ is  linearly dependent upon  the  initial condition
 $(\bar \varepsilon_0, \hat \varepsilon_0)$,  and  hence for suitably small $|(\bar \varepsilon_0,\hat \varepsilon_0)|$,   we  have $\hat e_A(t)   +  \hat e_B(t)  >0$.
Thus   $\hat e(t) = \hat e_A(t)   +\hat  e_B(t)  \ra  0$ as $t \ra \infty$  without changing sign in  any component.
\endproof

\begin{remark}
It is worth  considering  the  sense  in  which the  multi-agent Problems  \ref{P1} and  \ref{P2}  have   been solved with a distributed  control system: what information  and assumptions are required to hold  globally (for all  agents), and which ones are  local (information that only  needs to be known by individual agents)?
The information that must be available for the  purpose of controller design is as follows:
\bn
\item    All  agents  require knowledge of the exosystem  dynamics $S$,  however only  the  informed agents are able to  directly detect  the states  of the exosystem. The uninformed agents  detect  the state of the exosystem using  information obtained from  the  informed agents,  via  the  communication  network.
\item The control law (15) requires the design of the feedback matrix $F_i$ and the feedforward matrix $G_i$ for each agent.
$F_i$ requires knowledge of the plant dynamics $(A_i,B_i,C_i)$,  and  estimates of the initial  states $x_{i,0}$ and $w_0$ of the  $i$-th  plant and the exosystem, while  $G_i$ requires solutions to matrix equations (\ref{Pi1})-(\ref{CPi2}).  Thus the design of these  matrices can be done  locally, provided  $S$ is available. 
    \item  For the  informed agents $i \in \{1, \dots, l\}$,  design of the observer gains $L_{1,i}$, $L_{2,i}$ defined in (21)   can be also be done locally,  however  there  must  be  an agreement among the controller designers on the values of $\lambda_0$ and  $\mu_0$ to  be  used.
 \item   For the  uninformed agents $i \in \{l+1, \dots, N\}$, design of the observer gains $L_{i}$   defined in (21) can  be done  locally, provided all  these agents have  knowledge  of the parameters $\lambda_0$  and  $\mu_0$. Additionally,  the  controller design procedure for these agents requires knowledge of the Laplacian submatrix $\mathcal{L}_{33}$ so that  suitable $\gamma$ satisfying (22) can  be selected.
     \en
     Thus the controller design method  of \cite{SH12b} requires global knowledge of the exosystem  dynamics $S$. Assumptions  of  this kind are widely used  in   problems of multi-agent consensus tracking control,  for example   \cite{HGCH}-\cite{SH12a} among  many others.   In  Section V,   we provide some further discussion on  the cooperative nature of the controller design method in  the context  of an  aircraft control example.
     \end{remark} 

\section{Example}
\label{secex}

\begin{figure}
		\centering
		\includegraphics[scale=1]{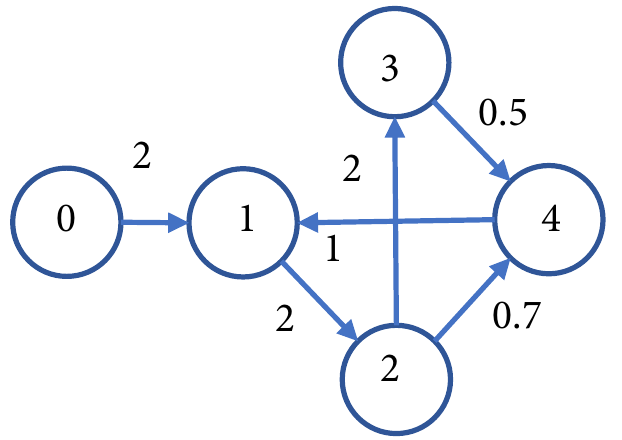}
		\caption{Network of four interconnected aircraft}
		\label{fig:topA}
	\end{figure}

In order to show the effectiveness of our proposed method, we adopt an example from \cite{YW13}. In this example, we consider four networked research aircraft known as MuPAL-$\alpha$ connected as shown in Fig.~\ref{fig:topA}. It is desired for the aircraft to track a given sideways velocity and a given roll angle. 
 The exosystem states are defined as $w= (r_v, r_{\phi 1 }, r_{\phi 2}, d_{\phi1},d_{\phi 2})^T$, where $r_v, r_{\phi 1 }, r_{\phi 2}$ are the states of the reference signal, and $d_{\phi1},d_{\phi 2}$ denote the sensor noise in the channel of roll angle. The matrix $S$ is defined as follows:
\bes
S=\bbmat 0 &0& 0& 0& 0\\
0& 0 &-\frac{2}{3}& 0 &-0.1\\
0& \frac{1}{4}& 0 &0 &0 \\
0& 0& 0& 0& 1\\
0& 0& 0& -1& 0
\ebmat
\ees
The states of each aircraft are considered as $x_i=(v_i,r_{ri}, \phi_i, y_{ri}, T_{ai}, T_{ri})^T$, the  sideways velocity, roll rate, roll angle, yaw rate,  and delays of the two commands, respectively.   The measured output $y_i$ is considered as $y_i=(v_i,\phi_i)^T$,  and  $u_i=(\delta_{a_ci},\delta_{r_ci})^T$,  the aileron deflection  and rudder deflection commands.
The regulated outputs $e_i$ are the  tracking errors of sideways velocity and roll angle. 

The state matrix of each aircraft for $i=1, \ldots,4$ is given as
\bes
A_i\ns=\ns\bbmat
-0.178 \ns&\ns 6.079\ns&\ns 9.763\ns&\ns -65.623 \ns&\ns0\ns&\ns 2.890\ns\\
-0.057 \ns&\ns-3.810\ns&\ns 0 \ns&\ns1.343 \ns&\ns-10.750 \ns&\ns1.187\ns\\
0 \ns&\ns1.000\ns&\ns 0\ns&\ns 0.094\ns&\ns 0\ns&\ns 0\ns\\
0.025\ns&\ns -0.062\ns&\ns 0 \ns&\ns-0.475\ns&\ns 0.345\ns&\ns -2.220\ns\\
0\ns&\ns 0\ns&\ns 0\ns&\ns 0\ns&\ns -11.111\ns&\ns 0\ns\\
0\ns&\ns 0\ns&\ns 0 \ns&\ns0 \ns&\ns0 \ns&\ns-11.111
\ns\ebmat
\ees
Also, for $ i=1, \ldots,4$ we have
\bes
\bal
B_i=&\bbmat  0 &-2.8900\\
10.7500& -1.1870\\
0& 0\\
-0.3450& 2.2200\\
22.2222& 0\\
0 &22.2222 \ebmat, \
C_{y,i}   =\bbmat 1 &0&0&0&0&0\\
                   0& 0& 1& 0 &0& 0\ebmat\\
H_{e,i}=&\bbmat -1 &0&0&0&0\\
                        0& 0 & -1 &0& 0\ebmat, \
H_{y,1}=\bbmat 1 & -1& 0&0&0\\
                        1 & -1& 0& 0 & 0\ebmat
                        \eal
 \ees
Also $C _{y,i}  = C_{e,i} $  and  $ H_{y,2}=  H_{y,3}=  H_{y,4}= 0$.
Conditions (A.1)-(A.6) may  readily be checked to  be valid;  in solving (\ref{Pi1})-(\ref{CPi2}) we used
\bq
\Gamma \ns & =  & \ns \bbmat    -0.0045 &   -0.0877  &   0.0472 &   -0.0145  &    0.0327 \\
                                               0.0112 &   -0.0427  &  -0.0139  &   -0.0065  &   0.0100
                                               \ebmat \nn \\
\Pi   \ns & = & \ns  \bbmat     1.0000   &       0      &    0             &         0      &       0  \\
                                      0.0002   &  0.2480  &  -0.0138  &    0.0000  &   -0.0866 \\
                                        0          &        0      &   1.0000   &        0        &        0  \\
                                     -0.0022  &  0.0211  &   0.1467   &  -0.0002   &   -0.0076 \\
                                     -0.0089  &  -0.1773   &  0.0837  &    -0.0231  &     0.0659 \\
                                       0.0223 &   -0.0847 &   -0.0329  &   -0.011   &     0.0203
                        \ebmat \nn
    \eq  
  
 From the network  graph in Figure \ref{fig:topA},  we see that the  informed group of  agents  consists  of agent 1,  while agents 2,  3 and 4 are  uninformed. The Laplacian matrix of the digraph is
\bes
\mathcal L = 
\bbmat
0& 0 &0 &0& 0\\
-2 &3 &0 &0& -1\\
0 &-2 &2& 0& 0\\
0& 0 &-2& 2& 0\\
0& 0 &-0.7 &-0.5& 1.2
\ebmat
\ees

The  distributed  nature of the control  scheme can    be understood  in terms of this  aircraft example system.  The controller design of the  matrices  required  for the control  scheme (\ref{ulaw1})-(\ref{ulaw3})  for each aircraft can be done without knowing the identity (flight dynamics) of the  other aircraft in the network, provided there  is a consensus on  the location of  closed-loop poles.  Regarding knowledge of the  communication  digraph $\mathcal G$, aircraft in the informed group need only know of  those aircraft to  whom  they are directly linked by an  edge of  the digraph. Aircraft  in the uninformed group  require sufficient information  to  enable them to compute  the  Laplacian  submatrix $\mathcal{L}_{33}$.
The exosystem  (\ref{exos}) represents a flight maneuver that all the aircraft are to  execute.  The maneuver involves varying the sideways velocity and  roll angle of each  aircraft.  Global knowledge of  the $S$ matrix defining the exosystem dynamics means that all  aircraft  are aware of the  maneuver -  the  purpose of the control  scheme  presented  in [6] is to enable all aircraft in  the  network to {\em synchronise}  their execution of the maneuver with that of the leader aircraft. 

 The invariant zeros  of   each agent system $(A_i,B_i,C_{y,i})$ are at $\{ -50.54, \   11.11, \   11.11\}$.  Hence each  system  has one   minimum phase  zero at $-50.54$. There are 6  state   variables,  and two  inputs and outputs.  Thus (\ref{zerocond}) is satisfied,  indicating that a search  for feedback matrices to  ensure the state feedback $\tilde u = F \tilde x$  yields   a nonovershooting response on  the  nominal plant  (\ref{nomsysi}) is  likely to  succeed. It is  also worth noting that the system is  of  nonminimum phase,  due to the repeated right complex plane  zero at $11.11$.    \cite{SP11}  investigated the  transient response  of MIMO nonminimum phase systems, and found that,  although  overshoot could generally be avoided,  doing so often  came at the  cost of  undershoot,  and vice-versa.

\begin{figure}
		\centering
		\begin{tabular}{c}
		\includegraphics[width=8cm]{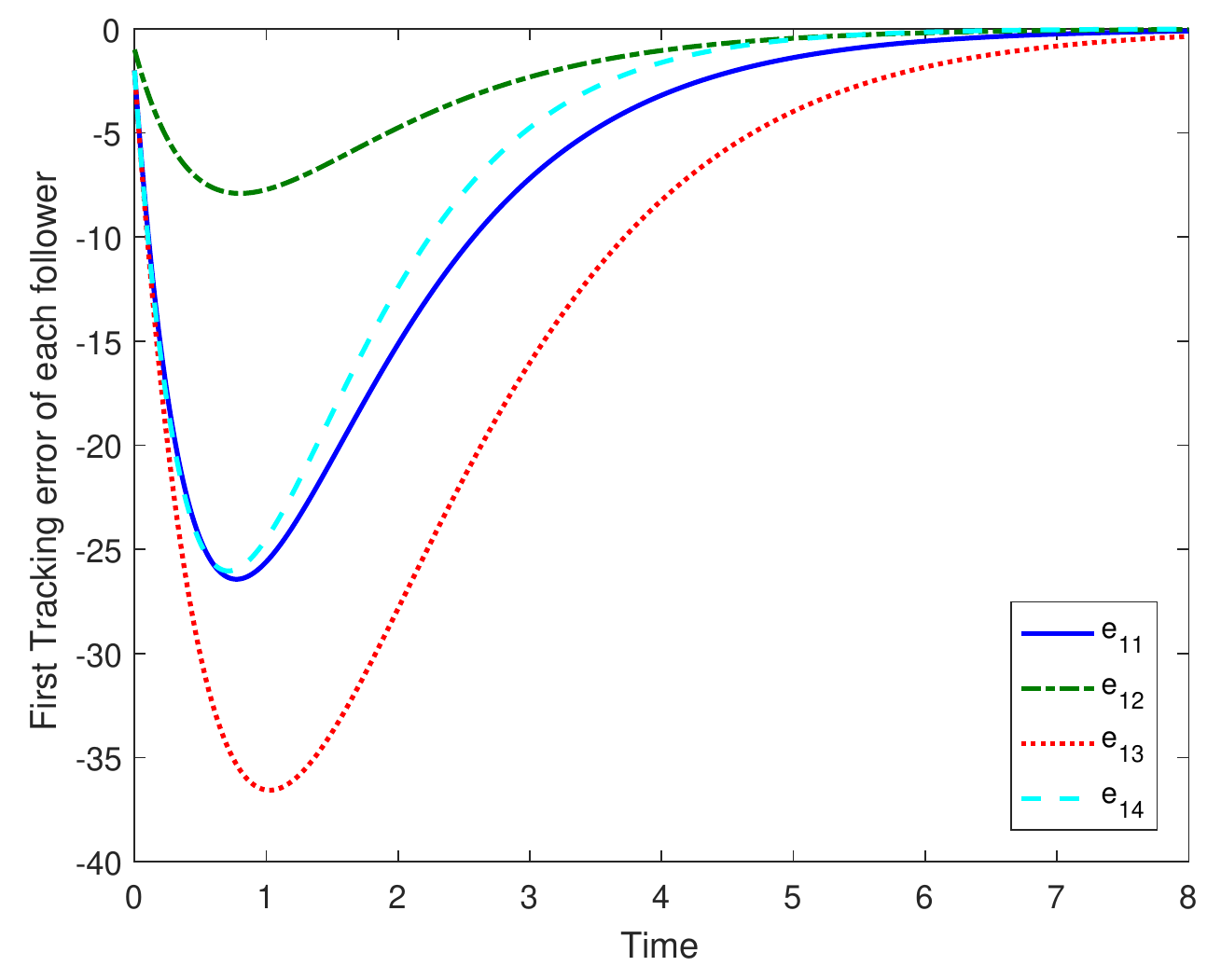} \\
		\includegraphics[width=8cm]{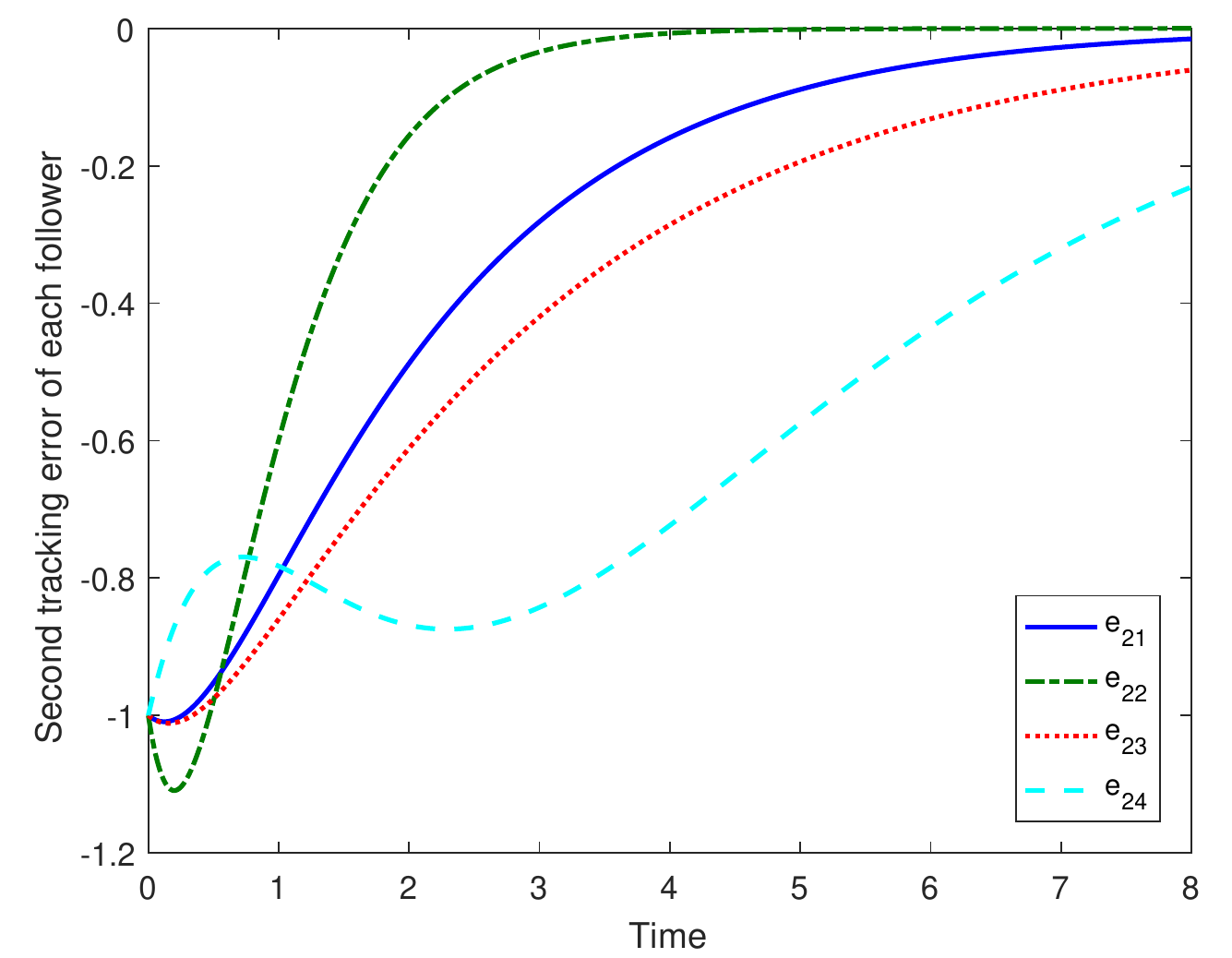}
		\end{tabular}		
		\caption{Tracking errors with no  error  in the initial state estimates}
		\label{fignoerror}
	\end{figure}

To  investigate the  application of the nonovershooting control method proposed in this  paper, we assume that estimates  of the  initial states of each agent and the exosystem  have  been obtained as follows
\bq
x_{1,0} & = &\bbmat -1 &    0   &   -1  &   1&     0  &   0\ebmat^T  \nn \\
x_{2,0} & = & \bbmat  0 &   -1   & -1  &   0  &  -1  &  -1  \ebmat^T          \nn  \\
x_{3,0} & = & \bbmat -1&     0 &   -1  &   1 &    0 &   -1 \ebmat^T \nn \\
x_{4,0} & = & \bbmat -1&    1&    -1  &   1 &    0&    -1  \ebmat^T  \nn \\
w_{0} & = &   \bbmat  1  &   1 &    0&     0&    0\ebmat^T \nn
\eq 
The \Nous toolbox \cite{PS12}  was used to seek such feedback matrices   for the  nominal  system (\ref{nomsysi}) of each agent, from initial conditions $\tilde x_{i,0}  =  x_{i,0}  - \Pi w_0$.   The toolbox  asks the user to nominate a desired  interval of the negative real line  for the location of the closed-loop eigenvalues. We chose the  interval $(-2.5, -.3)$,  and  in each case the search succeeded, yielding feedback matrices
\bq
F_1\ns & = &\ns \bbmat     0.006 &   0.025 &   -0.031 &   0.120 &   0.631 &  -0.250 \\
                                0.061  &  0.023 &  -0.354  &    1.08    &   1.38 &  -2.04
                                           \ebmat \nn \\
F_2\ns &  = & \ns \bbmat     0.005 &    0.015 &   -0.025  &     0.158  &     0.578   &    -0.250  \\
                                 0.055  &    0.004  &    -0.387  &    1.29  &     1.39 &   -2.07
                                        \ebmat  \nn \\
F_3 \ns  & =  & \ns \bbmat     0.00  &    -0.007 &   -0.086    & 0.485 &   0.646  &   -0.230  \\
                                 0.004 &   -0.325 &    -0.978  &    5.056  &   1.61  &   -2.55
                                      \ebmat    \nn \\
 F_4 \ns & = &\ns   \bbmat          0.004  &    0.029 &    0.041 &    0.136   &  0.506 &    -0.239 \\
                                           0.056  &    0.049  &   -0.306  &   0.775  &   1.36  &    -2.00
                                \ebmat    \nn
  \eq 
   Figure \ref{fignoerror}  shows the tracking errors for the  sideways velocity and  roll  angle  for each agent,  when the
 the control law  $\tilde u_i = F_i \tilde x$  is applied to  the nominal plant (\ref{nomsysi}) with initial  condition $\tilde x_{i,0} = x_{i,0} -\Pi_i w_0 $.   These  yield nonovershooting  tracking   errors  $\tilde e_i$ for both outputs of all agents. This  situation  corresponds to the  tracking errors  that would be  observed from the   multi-agent system (\ref{eqsys})  under the distributed  dynamic output feedback  controller  (\ref{ulaw1})-(\ref{ulaw3}) if there were  no error in the estimates of the initial agent and exosystem  states, and then   $\bar \epsilon(0) = 0$ and $\hat \epsilon(0) = 0$.  
	
 To implement the dynamic controller (\ref{ulaw1})-(\ref{ulaw3}),  we chose  $\mu_0 = -12$ and  $\gamma = 24$. These choices satisfy (\ref{gammadef}) as $\rho(S)$  lies on the  imaginary axis, and
$\rho(\mathcal L_{33}) = \{ 1.2, \   2,  \     2\}$.   Observer  gain matrices  $L_{1,1}$, $L_{2,1}$ and $L_i$ for  $i \in \{2,3,4\}$ to  ensure the closed-loop matrices in (\ref{Acci}) have spectrum  lying to the  left of $\mu_0$
were obtained using  the MATLAB$^{\textrm{\tiny{\textregistered}}}$  \place command:
\be
L_{11}  =      10^6\bbmat
   -1.004  &  2.628  \\
   -0.020   & 0.051\\
   -1.004   & 2.628\\
   -0.153   & 0.401\\
   -0.000   & 0.000\\
    0.000   & -0.000 \ebmat,
 L_{21}     =      10^7\bbmat
    0.593  & -1.556\\
    0.492   &-1.294\\
   -0.108   & 0.284\\
    0.427   & -1.109\\
    0.402    & -1.058
    \ebmat    \nn
   \ee
   \vspace{-.2cm}
      \be
L_2 = L_3 = L_4   =
\bbmat
  -32.5424  &   -3.1068  \\
   -1.8177   & -147.3808 \\
   -0.1349  & -27.7723  \\
    3.7878  &  -17.6487  \\
    0.1207   &   1.7550  \\
   -0.2203   &  0.4804
   \ebmat   \nn
   \ee 
     \vspace{-.2cm}

\begin{figure}
		\centering
		\begin{tabular}{c}
		\includegraphics[width=8cm]{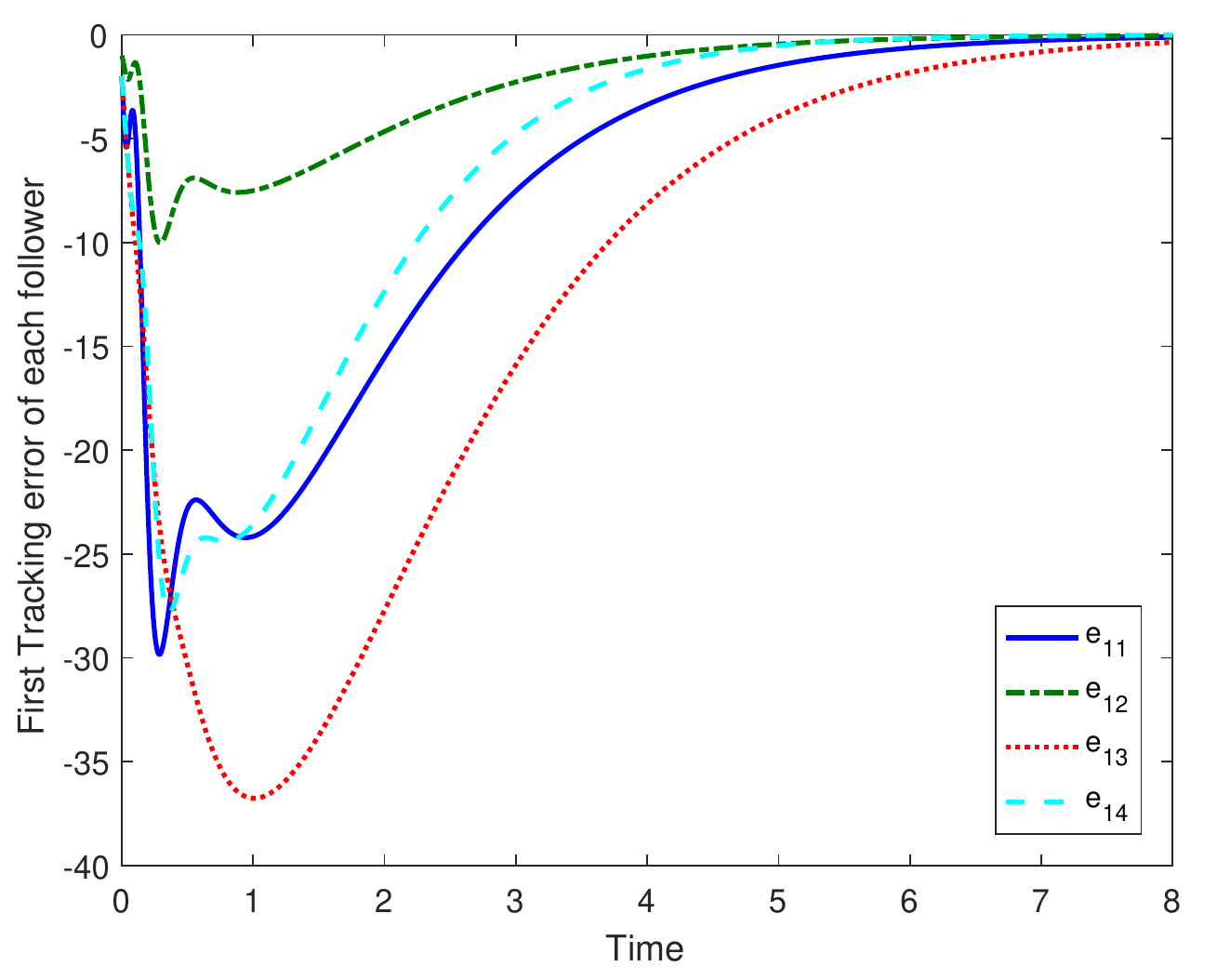} \\
		\includegraphics[width=8cm]{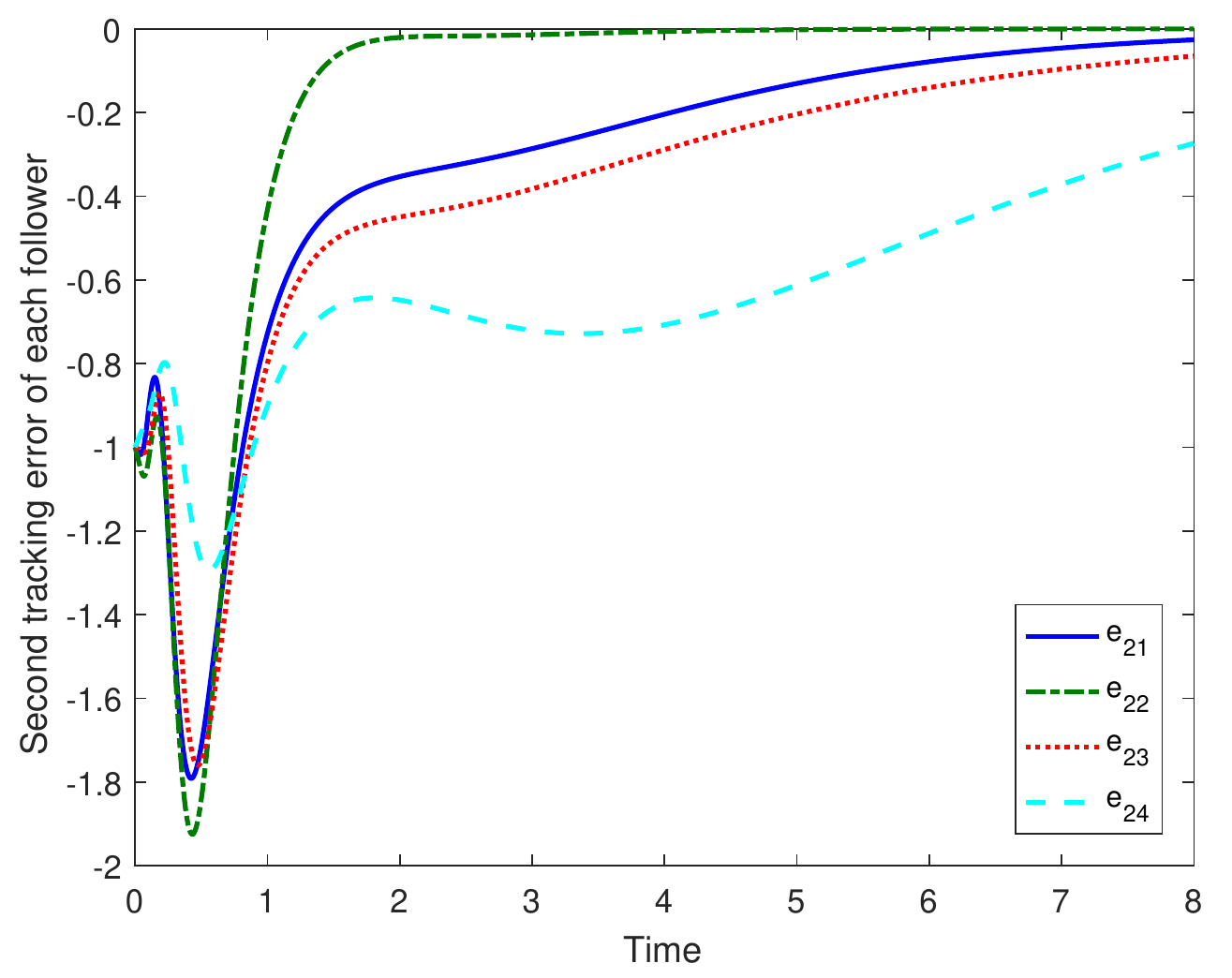}
		\end{tabular}		
		\caption{Tracking errors allowing for errors in the initial state estimates}
		\label{figwitherror}
	\end{figure}

 To  show the effect of the initial state  estimate errors $\bar \epsilon(0)$ and $\hat \epsilon(0) $ in the system response, we  shall assume  these errors are 1\% of the state estimates. Hence the initial  states of systems  (\ref{clinfb}) and  (\ref{clnoninfb})  are
\bq
\bar \xi(0)   & = & 1.01 \bar x(0),  \ \bar \eta(0)   = 1.01  \bar w(0)  \nn  \\
\hat \xi(0)   & = & 1.01 \hat x(0), \ \hat  \eta(0)   =  1.01 \hat  w(0) \nn
\eq 		
	 Figure \ref{figwitherror} shows the tracking errors for the  sideways velocity and  roll  angle  for each agent,  assuming these errors  in the estimates of the initial  states.
	 We observe  that both tracking errors from all  four agents converge to zero  without changing sign,  and thus  overshoot is avoided in both outputs of all four  agents  -   a total of 8 outputs. If the  dynamic controller (\ref{ulaw1})-(\ref{ulaw3})  had  been  designed  using the  methods  of \cite{SH12b} for the choice of the state feedback matrices,  then  the tracking errors would  also converge to  zero,  however the transient responses of each  output component would  be expected to  involve some  overshoot,  as  may be  observed in Figure 3 of \cite{SH12b}.  Overshoot would  occur even if the  initial  state estimation  errors  were zero \cite{SH12a}.

The additional contribution of the control methods in  \cite{SN10}  is to  choose the feedback matrices in  a manner that avoids  overshoot   and  hence enables  the  transient  period  of the control  action -  during which synchronisation is being achieved and when  the  aircraft to  do  not yet all  have  the same sideways velocity - to be conducted  in a smoother and potentially  less dangerous  manner. 

\section{Conclusion}
We have  investigated the  problem  of designing a consensus control  scheme to solve the  output regulation  problem  with a desirable transient response  for a family of  linear  multi-agent systems.  The  distributed consensus output regulation  scheme of Su  and  Huang  was  combined with the nonovershooting feedback  control  scheme of  Schmid and  Ntogramatzidis  to  achieve output regulation  without overshoot for all  agents,  under Assumptions (A.1)-(A.8). The author's believe this to  be the first control methodology to  achieve  a  nonovershooting transient response for MIMO  multi-agent consensus problems. 

Theorem \ref{mainthm}  guarantees the existence  of  a  neighbourhood  of  the estimated initial state
$ \tilde  x_{i,0}  $ such that,  if the actual  system initial state lies  within this  neighbourhood, then nonovershooting output regulation  will  be achieved by the distributed dynamic measurement output feedback  controller in (\ref{ulaw1})-(\ref{ulaw3}).   Estimating the  size of this neighbourhood is  an open problem, however, the  neighbourhood  can be adjusted by the choice of $\lambda_0$,  $\mu_0$ and  $\gamma$ in (\ref{Acci})-(\ref{gammadef}).  The neighbourhood becomes larger  if  the  initial states  of some agents are known,  and  also if  the  nonovershooting  behaviour  is  only required in a  selection of the agent outputs.   In practice, the size  of this  neighbourhood can be estimated with the assistance of the  \Nous MATLAB$^{\textrm{\tiny{\textregistered}}}$  toolbox \cite{PS12}. This toolbox allows the user to obtain state feedback matrices for a nonovershooting response from  the estimated system initial state, for each agents.  Combining these within (\ref{ulaw1})-(\ref{ulaw3}) and simulating the response of (\ref{eqsys}) from a range of error estimates of the initial system state will enable this neighbourhood to be approximated.

\section{Acknowledgements}
The authors thank  the Associate Editor and the anonymous reviewers for their  extensive and  insightful comments that have resulted in many improvements to the paper. 

\label{secconc}

\section{Appendix}
\underline{Proof of Lemma \ref{lem63}}:
Assume  firstly that $g(t) >0$ for all $t \geq 0$. Define  $\lambda = \min\{\lambda_i:    i \in \{1, \dots, m\}\}$.  Then $ \mu < \lambda$ by assumption.
Define  $f_1: \real \rightarrow \real $ with
\be
f_1(t) =  - \sum_{i=1}^n \ e^{\mu_i t} (|\alpha_i | + | \beta_i|)
\ee
Then $f_1(t) \leq f(t)$ for all $t \geq 0$.
As $f_1$ and $g$ are the sums of finitely many  negative real exponential functions, they  have  finitely  many  local  extrema, and there  exists   a $\bar t >0$ such that both  $f_1$ and $g$ are monotonic on the interval $t \geq \bar t$,  and  $f_1(t) \rightarrow  0$ and  $g(t)\rightarrow 0$ as $ t \rightarrow  \infty$.
Hence we  have $\delta_1 > 0$ such that
\be
\frac{1}{\delta_1} = \sup\left\{\frac{-f_1(t)}{g(t)} : 0 \leq t \leq \bar  t\right\}
\ee
and so  $ 0 < g(t) + \delta_1f_1(t)  $ for all $ 0 \leq t \leq \tilde t$.
Consider    $t >  \bar t$. As   $f_1$ is a SEDS  function with  rate $\mu$, we know that for $t > \bar t$
\be \label{fineq}
-f_1(t)  \leq  |f_1(\bar t)| e^{\mu (t-\bar t)}
\ee
Assume  without  loss of generality that  the   $\{\lambda_1, \dots, \lambda_m\}$ are  ordered so  that  $\lambda_1 < \lambda_2  < \dots < \lambda_m$. Then $\beta_m e^{\lambda_m t}$ is the dominant term of  $g$ as  $t \ra \infty$.  Also the assumption that $g(t) >0$ for all $t \geq 0$ implies  $\beta_m >0$. We next introduce  the set of integers $T_1 = \{i \in \{1,\dots, m\}: \beta_i \beta_m >0\}$ and the  exponential  function
\be g_1(t) = \sum_{i \in T_1} \ \beta_i e^{\lambda_i t }
\ee
Clearly  $g_1(t)  >0$ for all  $t \geq 0$,  and as   $m \in T_1$, we  see that  $\beta_m e^{\lambda_m t} $ is the dominant term of   $g_1$.
Hence we can introduce the function  $\gamma(t) $ such that $g(t) =  \gamma(t)g_1(t)$. Then $ 0 < \gamma(t) \leq  1$ for all $  t \geq \bar t$, and  $\gamma(t) \ra  1$ as $t \ra \infty$.  Define $\gamma_0  > 0$    with $\gamma_0 = \inf\{\gamma(t): t \geq \bar t\}$;  we then have  for  all $t  >  \bar t$ that
\be \label{gineq}
g(t) \geq  \gamma_0 g_1(t)  \geq\gamma_0  g_1(\bar t) e^{\lambda_1(t - \bar t)}
\ee
From (\ref{fineq}) and (\ref{gineq}),  we  obtain for $t > \bar t$,
\bq
\frac{-f_1(t)}{g(t)}
& \leq  &  \frac{ |f_1(\bar t)| e^{\mu (t-\bar t)}} { \gamma_0 g_1(\bar t) e^{\lambda_1(t - \bar t)} } \\
& <   &  \frac{ |f_1(\bar t)|}{  \gamma_0 g_1(\bar t)  }
\eq
as  $\mu < \lambda_1$. Defining $\delta_2 =    \frac{ \gamma_0 g_1(\bar t) }{    |f_1(\bar t)| }  >0 $,  we  obtain
$ 0 < g(t) + \delta_2f_1(t)  $ for all $ t  > \bar  t$.  Finally   choosing  $ \delta  = \min\{\delta_1,\delta_2\}$, and noting that $f_1(t) \leq f(t)$,  we  have   $g(t) +  \delta f(t) >0 $ for all $ t \geq 0$.  A  similar argument  can be  used  if $g(t) < 0$ for all $t \geq 0$,   and the result follows.
\endproof


\end{document}